# Modeling of disordered materials: radial distribution function vs. vibrational spectra as a protocol for validation

F. Gaspari (1,[*]), I.M. Kupchak (1,2), A.I. Shkrebtii (1), J. M. Perz (1)

(1) University of Ontario Institute of Technology, Oshawa, ON, L1H 7L7, Canada
(2) V. Lashkarev Institute of Semiconductor Physics NAS, Kiev, 03028, Ukraine

ABSTRACT

As molecular dynamics is increasingly used to characterize non-crystalline materials, it is crucial to verify that the numerical model is accurate enough, consistent with experimental data and can be used to extract various characteristics of disordered systems. In most cases the only derived property used to test the "realism" of the models has been the radial distribution function. We report extensive ab-initio simulation of hydrogenated amorphous silicon that demonstrates that although agreement with the RDF is a necessary requirement, this protocol is insufficient for the validation of a model. We prove that the derivation of vibrational spectra is a more efficient and valid protocol to ensure the reproducibility of macroscopic experimental features.

Important macroscopic properties of non-crystalline systems include radial distribution function (RDF), bond angle distribution, vibrational spectra, density of electronic states and optical (or mobility) gap. Hydrogenated amorphous silicon (a-Si:H), a prominent disordered material, has been the subject of intensive investigation for at least 30 years. An extensive literature, covering most important a-Si:H properties[1,2], provides a good test for the verification of the validity of a particular model. Due to its importance in photovoltaics [2,3] and microelectronics [4], many theoretical techniques have been employed to establish a realistic microscopic model of a-Si:H, including classical [5,6], density-functional-based tight-binding molecular dynamics [7], and *ab-initio* molecular dynamics (AIMD) [8,9]. Some of these simulations have been used recently to investigate important processes in a-Si:H, including hydrogen dynamics and the Staebler-Wronski effect [10]. However, verification of the "realism" of the models has mostly been limited to the well established procedure of derivation of the RDF and its comparison with experimental data. We will demonstrate, however, that RDF in fact cannot be used as universal parameter to verify the accuracy of numerical approach or particular model in general.

---

[*] Corresponding author, email: franco.gaspari@uoit.ca

For instance, Su & Pantelides [9] use AIMD to simulate a-Si:H and to analyze H diffusion, employing the Radial Distribution Function (RDF) to validate the model. This results in an incorrect assignment of the frequency of an H atom hopping between two Si atoms to the experimentally observed vibrations in a 2000-2100 cm$^{-1}$ frequency range, the most significant hydrogen vibrational signature in amorphous silicon (a more detailed discussion on this feature will follow). Nevertheless, agreement of the radial distribution function continues to be the "standard" protocol for judging the correctness of models not only of a-Si:H, but of disordered materials in general, as indicated below.

Tuttle [11] calculates the properties of hydrogen in crystalline silicon and in a realistic model of hydrogenated amorphous silicon by using Density Functional Theory. The *a*-Si:H model employed in this work was originally developed by Guttman and Fong [12]. The "realism" is verified by bond length and RDF.

The choice of using RDF also for other disordered structures is shown by Monaco *et al.* [13] in experimental studies of the effects of densification on the vibrational dynamics of $Na_2FeSi_3O_8$ glass, and by Masciovecchio *et al.* [14], in experimental studies of $SiO_2$. Munejiri *et al.* [15] present first-principle Molecular Dynamics simulations of liquid Ge, and use RDF as a validation parameter. Akola & Jones [16] use combined Density Functional/Molecular Dynamics simulation to study liquid and amorphous structures of $Ge_{0.15}Te_{0.85}$, and GeTe alloys. Again, they use RDF as the link to the experimental results. Finally, Urli *et al.* [17] use tight-binding MD to study point defects in pure amorphous silica. It is interesting to note that the same authors state "…the atomic structure of these (amorphous) materials has not yet been completely resolved and the analogy with the Continuous Random Network (CRN) has not been fully established. In particular, there exists no evidence that the coordination number of the prototypical a-Si is equal to 4, even though computer models do indicate that this might very well be the case; experiment has not yet been able to provide a definite answer to this question…"

If the goal of future work in the field is to expand the investigation of the properties of disordered materials, by examining more complex processes and characteristics such as the Staebler-Wronski effect, the density of states, atom migration dynamics, optical absorption, conductivity, *etc.*, then one must ask the question whether these properties are being examined or extracted using a realistic structure. What defines a realistic structure? Although agreement with the RDF is a necessary requirement, is this protocol sufficient for the validation of a numerical model?

Experimental measurements of RDF are not common and difficult to implement; furthermore, no recent experimental data are available for a-Si:H. In contrast, vibrational experiments are essentially more common, accurate and, most notably, linked to various processes in disordered materials [18,19]. Therefore, apart from providing a useful means to access microscopic properties of non-crystalline materials, the verification of the validity of theoretical models can be based on calculated vibrational spectra. However, there is no satisfactory theoretical formalism available to decode vibrational spectra in terms of atomic level processes within non-crystalline materials. Standard theoretical vibrational techniques (see, [20] and refs. therein) are not efficient for non-crystalline network due to lack of translational symmetry and complicated atomic dynamics.

To overcome these problems, the authors have developed a comprehensive approach, that combine *ab initio* molecular dynamics and improved signal processing, which has proven to be a powerful tool for the determination of a number of fundamental properties of a-Si:H [21,22]. In brief, a variety of numerical samples using a 64 Si atom supercell with a wide range of hydrogen concentration was simulated by AIMD, and visualized atom dynamics was used to extract a-Si:H frequencies by the signal analysis method MUSIC [23]. We have found, however, that the vibrational spectra of the amorphous 64 Si supercell with H atoms was practically impossible to analyze due to excessive number of frequencies (compared, e.g., to the crystalline Si). Most importantly, hydrogen and silicon diffusion as well as H-bond switching modify the atomic vibrations during the long MD run, thus further complicating vibrational analysis. To overcome this difficulty, we have developed and extensively tested an approach that combines the signal analysis method MUSIC and Fourier transform [21]. This paper, by further developing the above numerical methods, addresses a fundamental issue of verification of model validity in general by focusing on the correlation of non-crystalline material structures with vibrational properties and microscopic bonding. Using a-Si:H as an example, we were able to clearly separate all vibrational modes of the Si-H complexes and their *real time* genesis, and uncover various types of a-Si:H instabilities. In particular, we have observed and correlated the signals at 2000 cm$^{-1}$ and 2100 cm$^{-1}$ with their respective bond types, mono-hydrides and di-hydrides respectively [18]. Other characteristic frequencies, observed experimentally, have also been reproduced and correlated with their appropriate bonding structure.

In order to test the significance of a realistic reproduction of the RDF, we started with a crystalline silicon cubic supercell, containing 64 Si and 8 H atoms (a 128 Si supercell was used as well to prove that the 64 atom supercell is sufficiently accurate). The system was melted at 3000 K. After its equilibration in the molten state for 5 ps using Nosé thermostat [24], three different samples were created in the following ways: (a) the system was slowly cooled down to room temperature (RT) with cooling rate of $4.5 \cdot 10^{14}$ K/sec, and then relaxed to zero temperature (sample a); (b) the system was rapidly quenched to zero temperature (sample b); (c) the system was kept in the molten state for an additional 2 ps, and then quenched to zero temperature as well (sample c). Next, the atoms in all three systems were slightly distorted from their equilibrium positions in order to perform RT MD run for 10 ps to collect statistical data for vibrational analysis. The Nosé thermostat was off for the RT runs. It should be noted that samples (b) and (c) were obtained with a procedure non-conducive to a realistic structure, obtained experimentally.

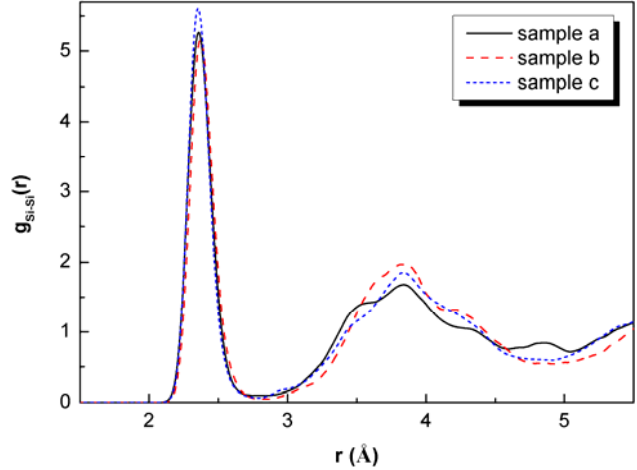

**Figure 1**. Radial Distribution Function for three samples prepared by three different methods from the same original (64 Si+8 H) supercell. Only sample (a) mimics the standard experimental cooling condition. Minor modifications of the RDF cannot be assigned to different structure or bonding.

In Fig. 1 we show the RDF for all three samples; strikingly there is very little variation among the three spectra. Moreover, RDF does not demonstrate features that can be interpreted in terms of microscopic processes in the system. However, when we compare the vibrational spectra for the same three cases (see Fig. 2), there is a dramatic difference between sample "a" (the "realistic" a-Si:H structure, as indicated by the strong signal at ~ 2000 cm$^{-1}$, the characteristic mono-hydride stretching mode) and samples "b" and "c", where the location and/or intensity of the various peaks do not reproduce the standard experimental observations [18,19].

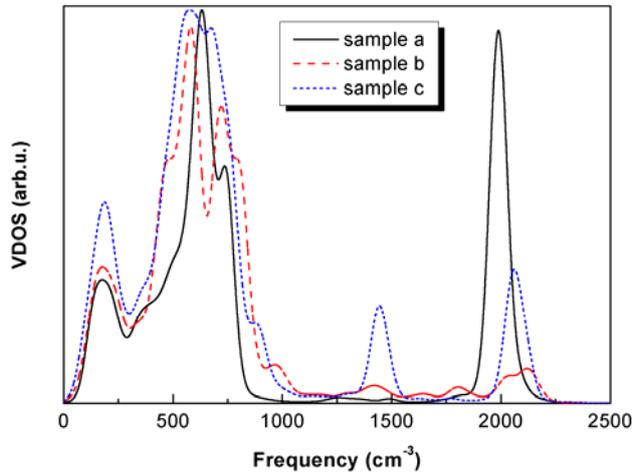

**Figure 2**. Vibrational Density of States (VDOS) for the same three cases shown in Figure 1. Note the essential modification of the spectra for the "unphysical" (b) and (c) samples. In particular, the main hydrogen stretching peaks are drastically reduced or practically disappear for (b) and (c) compared to properly prepared sample (a).

Figs. 3 and 4 show the RDF of the H–H pairs and the average bond angle distribution for the three cases, clearly indicating a structural difference among the three samples. Indeed, it is crucial that not only the underlying silicon network follow the amorphous structure, but that also the H-atoms be embedded in the proper way within the silicon matrix. This is a fundamental condition if the model has to be used to simulate processes connected with hydrogen dynamics, including the testing of some important models that have recently come to prominence as explanations of the Staebler-Wronski effect, such as the floating-bond model [25], and the hydrogen collision model [26].

The stability of the bonds is also important to verify of the accuracy and validity of a model. Fig. 5 (a, b, c) shows the intensity of the various characteristic vibrational frequencies in a-Si:H as a function of time for the three samples, as produced by our modified signal processing technique. The lighter color indicates a higher intensity, that is, the white region is where the peak is located. It is clear that we observe a stable bond only in case "a": the only fluctuations we observe for this case is in the intensity of the peak, however, both the 2000 cm$^{-1}$ and the 600 cm$^{-1}$ feature indicate very little shifts in the frequency position and a narrow peak (shown by the "spread" around the peak position). Furthermore, a lower intensity in the peak at 2000 cm$^{-1}$ corresponds to a higher intensity in the peak at 600 cm$^{-1}$, indicating that the fluctuations in peak intensities is due to a alternating predominance of the stretching modes vs. the rocking/bending/scissors modes. On the other hand, both sample "b" and sample "c" show large fluctuations in peak intensity, position and a large spread of the signal at 600 cm$^{-1}$.

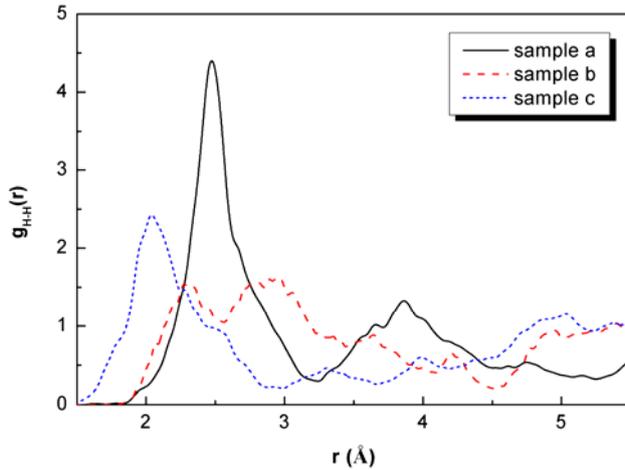

**Figure 3**. Radial Distribution Function for the H-H pairs for the three samples described in the text. Note the essential change of the distribution function due to difference in structure and bonding.

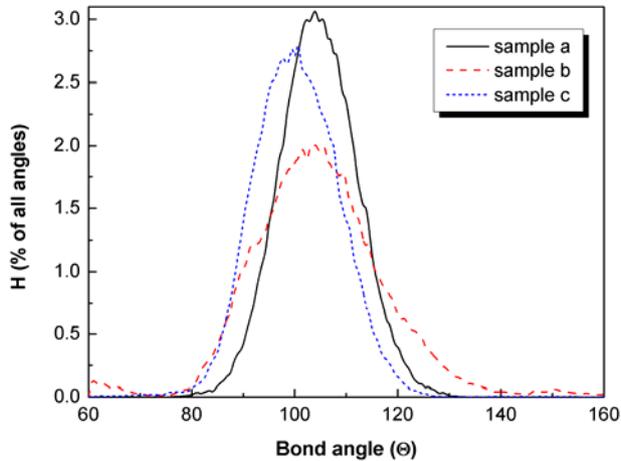

**Figure 4**. Variations of the average bond-angle for H-atoms for the same three a-Si:H samples as in the previous figures. The difference in the angle distribution is more pronounced than in the case of RDF, still it does not contain detailed information about microscopic properties of the samples.

A few different methods have been tried in the past to recreate a realistic a-Si:H, including removing Si atoms from a pure amorphous silicon network and "adding" H atoms [9,27]. It is clear, however, that in order to extract the relevant macroscopic parameters correlated with an amorphous structure, great care must be taken to verify the validity of the model beyond just the standard RDF of the Si network.

To confirm this we have run an additional test starting with a 64-atoms amorphous silicon sample with no hydrogen, obtained from the melt by slow cooling down to room temperature. We then added two hydrogen atoms at random positions within the unit cell, and the system was relaxed to zero temperature. These H atoms were placed in the vicinity of fourfold coordinated Si atoms, with all their bonds satisfied by other Si atoms, so that the creation of "free" interstitial hydrogen atoms seemed plausible. However, during the relaxation at zero temperature, both hydrogen atoms developed mono-hydride bonds with the nearest Si atoms, leading to a significant system distortion in the region of the mono-hydrides.

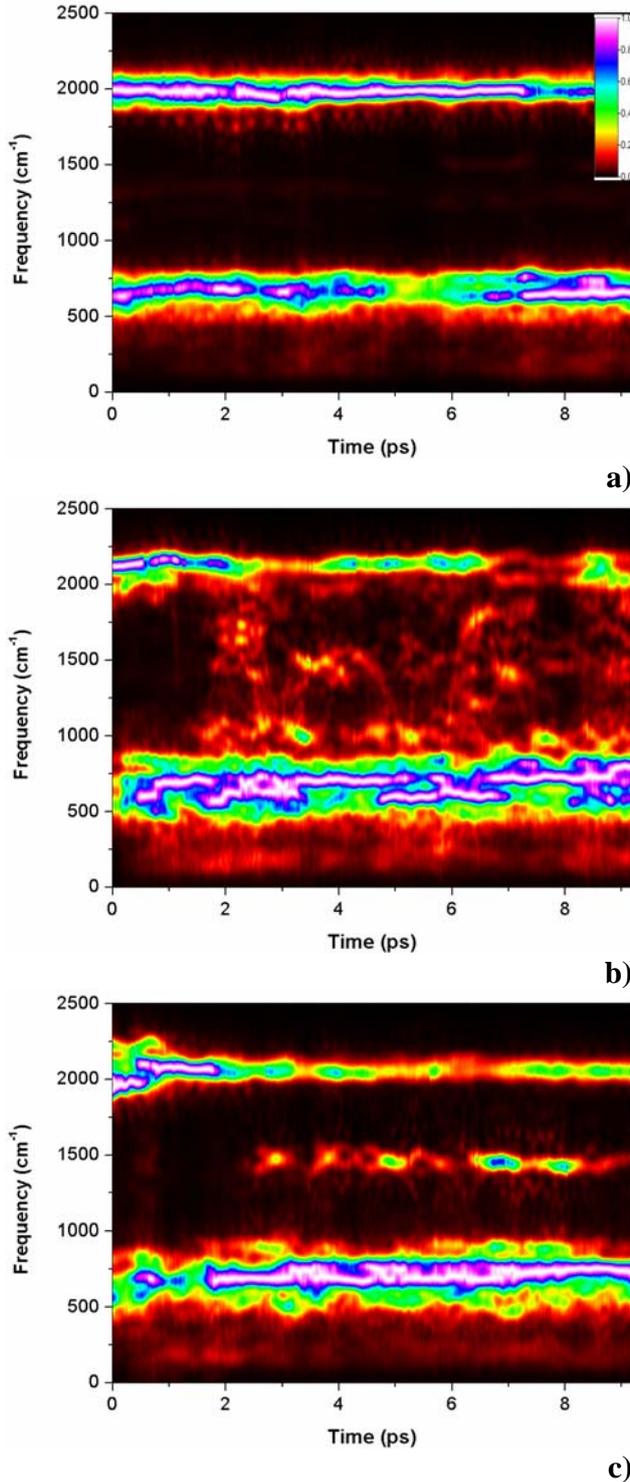

**Figure 5**. Real time dependence of the vibrational frequencies for the three samples: (a) top, (b) middle and (c) bottom. An intensity scale in shown in the inset on the graph for sample (a). Note the absence of the vibrations between two main low and high frequency modes in (a) and significant frequency instability for the samples (b) and (c).

Subsequently the atomic positions were slightly distorted from the relaxed positions, an MD run was performed at 300 K and statistical data were collected before annealing ("ba"). During the MD run, the system was stable – no essential changes occurred in the atomic structure. The system was then annealed at ~700 K. After annealing ("aa"), the system was relaxed to zero temperature again and an additional MD run was performed at 300 K to collect statistical data. During the annealing, one of the hydrogen atoms "jumped" to another Si atom, and this new mono-hydride bond remained stable for the rest of the simulation. Moreover, the "old" Si atom restored its bond with the nearest Si atom, which had been replaced by a mono-hydride before. The observations indicate essential structure and local bonding modification and to quantify this we extracted the RDF and the vibrational spectra before (ba) and after (aa) annealing. These spectra are shown in Figures 6 and 7. Once again, very little difference is observed between the two RDF spectra, while the vibrational spectra clearly indicate that only after annealing we were able to achieve a "realistic" a-Si:H sample.

We have compared a variety of a-Si:H samples with H content from 0 to 20% [21], and we have found that RDF practically does not depend on amount of hydrogen in the sample as well. Furthermore, all the calculated RDF agree reasonably well with the most recent and accurate RDF measurement for a-Si with no hydrogen [28]. This reflects the fact that the most probable distance between neighboring atoms is equal to a sum of the atoms' covalent radii. Even when hydrogen passivates the dangling bonds, this does not modify the Si–Si bond length. On the other hand, atomic vibrations do depend on microscopic bonding (bonds), their angular distribution, distortion or breaking. In fact, the experimental measurements [18,19,29] demonstrate a variety of spectral features that obviously require microscopic theoretical interpretation.

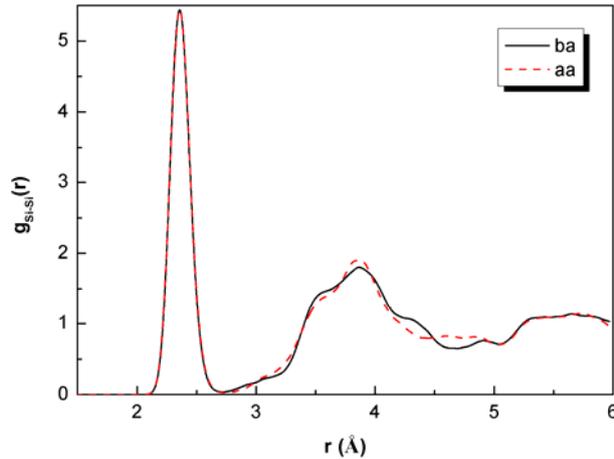

**Figure 6**. Radial Distribution Function for a test 64 (Si)+2 (H) system (there are two different structures but one system) before annealing (ba) and after annealing (aa). As for the case shown in Fig. 1, there are no noticeable differences in the RDF.

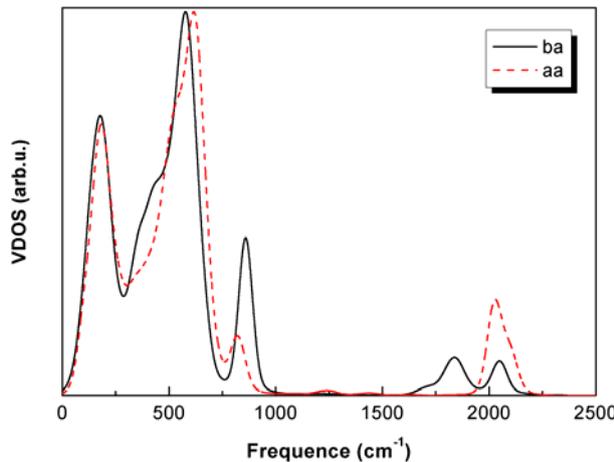

**Figure 7**. Vibrational Density of States for the 64 (Si)+2 (H) system before annealing (ba) and after annealing (aa).

In order to further verify the validity of our model, we have studied the special case of metastable Si-H-Si bonds. Darwich *et al*. [29], using infrared transmission spectroscopy (IRT) and infrared ellipsometry, observed a new, metastable feature at ~1730 cm$^{-1}$ during light-soaking, which they attributed to a three centre Si-H-Si bond (TCB). The authors remark that the theoretical calculation for the IRT frequency of such a complex had been a controversial issue, with the expected values located between 800 and 1950 cm$^{-1}$, and proposed that in fact the band at 1730 cm$^{-1}$ represents the stretching mode of the TCB. Indeed, we observe such a feature associated with a TCB type of bond between 1500 and 1800 cm$^{-1}$ [21] from our simulations.

This behavior was also observed by Su and Pantelides [9] at high temperatures. The authors used AIMD to simulate the hydrogen migration process in a-Si:H. Their simulations were performed only at high temperatures (600 – 900 $^{O}$C) with a supercell of 64 silicon plus only 2 hydrogen atoms. The RDF was used to verify the agreement with experimental data. A typical occurrence observed by the above mentioned authors is the migration of a hydrogen atom from a silicon atom (A) to another silicon atom (B), with a calculated frequency between 2000 and 2100 cm$^{-1}$. These, however, are the characteristic frequencies of stable mono-hydride or di-hydride bonds rather than a metastable hydrogen bond.

Our AIMD results confirm Darwich's claim within experimental error. The decrease in the vibrational frequency with respect to that of a stable mono-hydride bond is due to the sharing of the hydrogen electron density between two Si atoms. This decreases the Si–H bond strength, increases the bond length and results in reduction of the vibrational frequency. Therefore, the band in the 1500-1800 cm$^{-1}$ region is the distinctive signature of all hydrogen metastable bonds, including the TCB bond, with variations in the frequency due to the different overlap between the H and the Si electron wave functions.

In conclusion, *ab-initio* molecular dynamics is a powerful tool for the modeling of complex systems, like solid state amorphous structures, and it is being used to probe fundamental properties and dynamics of disordered materials. However, in order to validate the simulation of complex structure, bonding, and diffusion, a protocol needs to be established for the verification of the "realism" of the simulated models. Using hydrogenated amorphous silicon as an example, we have unambiguously demonstrated that reproduction of the radial distribution function, used commonly in numerical simulations, is not sufficient and must be complemented with verification of other, more complex, macroscopic properties. Focusing on the vibrational modes of the amorphous system, we have proved that vibrational spectra represent a crucial testing tool for non-crystalline materials because of their complexity and sensitive link to structure and bonding configuration. Successful reproduction of all the experimentally observed vibrational features for a-Si:H proves the validity of our algorithm and indicates that hydrogen structure and dynamics are extremely sensitive to the parameters of the model. In order to correctly apply a numerical model to extract such important macroscopic parameters as density of states, optical gaps, and migration dynamics, the accuracy should be verified first by the derivation of the standard vibrational modes and comparison with experimental observation.

ACKNOWLEDGEMENTS

The research was supported by the Centre for Materials and Manufacturing/Ontario Centres of Excellence (OCE/CMM) "Sonus/PV Photovoltaic Highway Traffic Noise Barrier" project, Discovery Grants from the Natural Sciences and Engineering Research Council of Canada (NSERC) and the Shared Hierarchical Academic Research Computing Network (SHARCNET).